\documentclass[12pt,letterpaper]{article}
\pdfoutput=1
\usepackage{graphicx}
\usepackage{fullpage}

\usepackage{yfonts}
\usepackage{amsmath}
\usepackage{amssymb}
\usepackage{bm, bbm}      % bold math
\usepackage{color}
\def\pd{\partial} 
\def\=d{\, {\buildrel \rm def  \over =} \,}

\def\sqr#1#2{{\vcenter{\vbox{\hrule height.#2pt \hbox{\vrule width.#2pt height#1pt \kern#1pt \vrule width.#2pt}\hrule height.#2pt}}}}

\def\beq#1{\begin{equation} \label{#1}}
\def\eeq{\end{equation}}
\def\ben{\begin{equation*}}
\def\een{\end{equation*}}
\def\bequa{\begin{eqnarray}}
\def\eequa{\end{eqnarray}}

\def\Tr{\mathop{\mathrm{Tr}}}

\newcommand\eea{\end{eqnarray}}
\newcommand\bea{\begin{eqnarray}}

\def\beq{\begin{equation}}
\def\eeq{\end{equation}}

\def\H{{\mathbb H}}
\def\R{{\mathbb R}}

\title{Holographic Phases of R\'{e}nyi Entropies}
\author{
Alexandre Belin, Alexander Maloney, and Shunji Matsuura \\
\it Physics Department, McGill University, 3600 rue University, \\
Montr\'{e}al QC H3A 2T8,Canada
}

\begin{document}

\maketitle

\begin{abstract}
We consider R\'{e}nyi entropies $S_n = {1\over 1-n} \log  \Tr \rho^n$ of conformal field theories in flat space, with the entangling surface being a sphere.
The AdS/CFT correspondence relates this R\'{e}nyi entropy to that of a black hole with hyperbolic horizon; as the R\'{e}nyi parameter $n$ increases the temperature of the black hole decreases.  If the CFT possesses a sufficiently low dimension scalar operator the black hole will be unstable to the development of hair.  Thus, as $n$ is varied, the R\'{e}nyi entropies will exhibit a phase transition at a critical value of $n$.  The location of the phase transition, along with the spectrum of the reduced density matrix $\rho$, depends on the dimension of the lowest dimension non-trivial scalar operator in the theory.
\end{abstract}

\newpage

\tableofcontents

\section{Introduction}

Entanglement entropies characterize the degree of entanglement present in a given quantum state, and in doing so probe  interesting features of strongly coupled quantum systems.
In conformal field theories, certain entanglement entropies are related to conformal anomaly coefficients ({see \cite{Calab-cardy-rev} for a review) and play an important role in conjectured holographic duals of conformal field theories \cite{RT}. 
In the condensed matter literature they have, for example,  been used to characterize topological order \cite{topological-order} and fractional quantum Hall edge states \cite{Li-Haldane}.  
%as well .
%For instance, in the seminal paper \cite{Li-Haldane}, Li and Haldane showed that the entanglement spectrum, the eigenvalue distribution of the reduced density matrix, of fractional quantum Hall effect contains the information of the counting of the edge conformal field theory.
 The goal of the present paper is to understand the extent to which entanglement entropies encode other interesting features of conformal field theories.

To define an entanglement entropy one starts by considering a quantum system which can be divided into two subsystems, $A$ and $B$, with associated Hilbert spaces ${\cal H}_A$ and ${\cal H}_B$.   The state of the system is given by a density operator $\rho_{AB}$ acting on the  tensor product Hilbert space ${\cal H}_A \otimes {\cal H}_B$.  The degree of entanglement between $A$ and $B$ is characterized by the reduced density matrix  $\rho = \Tr_{{\cal H}_B} \rho_{AB}$.  We are interested in basis independent quantities, so will consider the moments of the eigenvalue distribution of  $\rho$:
%, so we will consider the moments 
\bea
S_{n}={1\over 1-n}\log \Tr [\rho^n].
\label{Rnyi tr}
\eea
These are the R\'{e}nyi entropies, and $n$ is the R\'{e}nyi parameter.
In the limit $n\to 1$, $S_n$ becomes the entanglement (von Neumann) entropy $S_{EE} = -\Tr \rho \log \rho$. These 
R\'{e}nyi entropies have been considered in a variety of contexts \cite{renyi-original, Zyczkowski, calabrese-Lefevre, Steph-Rnyi-tran, Klebanov:2011uf,  Headrick:2010zt,Fursaev:2012mp}.

%constructed from this reduced 
%More recently, people have started to pay a lot of attention to reduced density matrices, the building block of the entanglement entropy.

%
%Superposition of states is a fundamental feature of quantum systems which never shows up in classical systems.
%When a state cannot be described by a single tensor product, it is call entangled.
%The entanglement entropy is used to quantify how much two subsystems are entangled.
%The physical importance of the entanglement has become clearer in the last decade.

%Our interest in this paper is to extract information about the reduced density matrix of certain conformal field theories.

%\adm{All of the dimensions of the spaces in the paragraph below were wrong, so I removed them}

The computation of the reduced density matrix $\rho$, and of the entanglement entropy $S_{EE}$, for an interacting quantum system is in general difficult.  We will consider the case of a $d$-dimensional CFT in flat space, %$\mathbb{R}^{d-1,1}$
 with $A$ and $B$ separated by a sphere. In this case conformal invariance relates the R\'{e}nyi entropy $S_n$ of the ground state to the thermodynamic entropy of the CFT on hyperbolic space $\H_{d-1}$ at temperature $T=T_0/n$ (as described in \cite{Casini:2011kv, Hung:2011nu}).  Here $T_0$ is a temperature related to the length scale of the hyperbolic space.  The computation of this thermodynamic entropy is still in general quite difficult.  We will therefore focus on CFTs with gravity duals, where the AdS/CFT correspondence relates this thermodynamic entropy to the entropy of a black hole in $d+1$ dimensional AdS space with hyperbolic horizon.  This allows us to compute  R\'{e}nyi entropies explicitly.

Solutions of Einstein gravity in AdS describing black holes with hyperbolic horizon are easy to construct; they are hyperbolic versions of the Schwarzchild solution with a cosmological constant.  However, black holes of this type are sometimes unstable \cite{Dias:2010ma}.  As the Hawking temperature $T\to 0$ (i.e. R\'{e}nyi parameter $n\to \infty$) they typically become more unstable.  Such an instability would lead to a non-analyticity in the R\'{e}nyi entropies, regarded as a function of $n$.  We find that this instability will occur for CFTs with a sufficiently low dimension scalar operator.  The reason is that a low dimension operator corresponds to a light scalar field in AdS.  If the field is too light, then at a finite, critical value of $n$ the scalar field will condense in the vicinity of the horizon.  The dominant solution is now a "hairy" black hole, with non-trivial scalar profile.\footnote{Similar phase transitions  occur in applications of AdS/CFT to condensed matter physics, most famously in the holographic superconductor \cite{Gubser:2008zu}.}

%A subtlety in the definition of Renyi entropies arises as the space on which the CFT lives has a singularity along the entangling surface, at which and one must define boundary conditions for the operators of the theory. \adm{this is incorrect} For example, one may demand that all fields are regular along the entangling surface. However, an alternative definition would be to allow operators to be sourced there. This will imply different conditions on the bulk scalars according to the AdS/CFT dictionary, namely defining if the bulk scalars should vary or simply be constant on the hyperboloid. We find that both definitions give rise to phase transitions in the Renyi entropies, but with somewhat different features. In the process of our analysis, we discover that hyperbolic black holes minimally coupled to a sufficiently light scalar field are unstable, even when the scalar is a normalizable mode of the hyperboloid. This instability, to the best of our knowledge, is new.

One advantage of this result is that it provides a clear dictionary relating properties of the R\'{e}nyi entropies (i.e. the eigenvalue distribution of the reduced density matrix) and natural CFT quantities.  For example, we will see that for a four dimensional CFT with a single scalar operator of dimension $\Delta < 3$ the second derivative of $S_n$ with respect to $n$ will be discontinuous at a value of $n$ which depends on $\Delta$.  We expect similar results for large $N$ CFTs with more complicated spectra of light operators.
It is interesting to note that a similar result was found in the $O(N)$ model using a direct field theory analysis \cite{sachdevON} (see also \cite{Steph-Rnyi-tran}).  In this case the R\'{e}nyi entropy was found to be non-analytic at $n = 7/4$.   Presumably our results are the bulk gravity version of this phase transition.\footnote{One subtlety is that the $O(N)$ model is apparently dual not to general relativity with a light scalar, but to a higher spin theory of gravity \cite{klebanovpolyakov}.   We expect that the transition at $n=7/4$ is related to an instability of a hyperbolic black hole in Vasiliev theory, but without a better understanding of the Vasiliev equations of motion it is difficult to make a precise statement .}

%\adm{fix up}
%For scalar fields which are above the Breitenlohner-Freedman (BF) bound, the phase transition happens for $n>1$.  Thus the R\'{e}nyi entropy is analytic in a neighbourhood of $n=1$, at which point it equals the entanglement entropy.  In this neighborhood the R\'{e}nyi entropy is computed by the usual Einstein black hole.  However, scalars which saturate the BF bound appear to lead to an instability infinitesimally close to $n=1$.  This may indicate that the replica trick -- which typically assumes analyticity of the R\'{e}nyi entropies in a neighbourhood of $n=1$ -- is invalid for these theories.  

Many approaches to entanglement entropy in quantum field theory -- such as the replica method -- implicitly assume that the R\'{e}nyi entropies are analytic in $n$.  We find that, for large $N$ field theories with a small gap, this assumption is not necessarily valid.  Thus the replica trick must be applied with care.  We note, however, that our phase transitions always occur  at values of $n$ which are strictly larger than 1.  Thus the R\'{e}nyi entropy is analytic in a neighbourhood of $n=1$, at which point it equals the entanglement entropy.  We do not see any indication of non-analyticity near $n=1$, except in rather exotic circumstances which we will comment on in section 4.

Our paper is organized as follows.  In section \ref{ent-reny-review}, we review R\'{e}nyi quantities and their computations  in quantum field theory.
In section \ref{HEE and HRE}, we review the holographic computation of the R\'{e}nyi entropy \cite{Hung:2011nu}.  We analyze the near horizon limit of the extremal black hole, which allows us to determine which black holes should be unstable in the $T\to 0$ limit.  This allows us to determine which conformal field theories will have R\'{e}nyi phase transitions.  

In the last two sections we will present numerical results which verify the existence of this phase transition.  In section 4 we begin with a simplified model (following closely work of \cite{Dias:2010ma}) in which the scalar mode which is constant on the hyperboloid condenses.  This is a particularly tractable case, as it preserves the hyperbolic symmetries.  This allows us to explicitly find the hairy black hole solution, which is used to compute R\'{e}nyi entropies and the spectrum of $\rho$.  However, the constant mode presented in this section is non-normalizable.  We therefore continue in section 5 to discuss normalizable modes.  Again, we demonstrate that the hyperbolic black holes of Einstein gravity are unstable, so that the R\'{e}nyi entropies will exhibit a phase transition.  The numerics are somewhat more difficult, as the modes no longer preserve the hyperbolic symmetry.  We therefore  present a linearized analysis -- which is sufficient to demonstrate that a phase transition will occur -- but leave the investigation of the hairy black hole for future work.

\section{R{\'e}nyi entropy in Conformal Field Theory}
\label{ent-reny-review}

%\subsection{R\'{e}nyi entropies and Hyperbolic Space}

%
%Suppose a quantum system has a Hilbert space $\mathcal{H}$. We divide the system into two subsystems $A$ and $B$.
%Each subsystem has Hilbert space $\mathcal{H}_{A}$ and $\mathcal{H}_{B}$ respectively.
%A state of the total system is described by a state vector $|\Psi_{AB}\rangle \in \mathcal{H}_{A}\otimes \mathcal{H}_{B}$.
%A reduced density matrix $\rho_A$ of the subsystem $A$ is defined by tracing over the Hilbert space $\mathcal{H}_{B}$ of the subsystem $B$, 
%\bea
%\rho_{A}={\Tr}_{\mathcal{H}_{B}}|\Psi_{AB}\rangle \langle \Psi_{AB}|.  
%\eea
In the following, we review the computation of R\'{e}nyi entropies for a relativistic (Lorentz invariant) quantum field theory.  
Consider a field $\psi$ in $d$ dimensional flat space.
The Euclidean signature metric is 
\beq
ds^2=dt^2+dx^2+d\vec{y}^2_{\bot}
%\cr
%&=&r^2d\theta+dr^2+d\vec{y}^2_{\bot},
\label{flat coord}
\eeq
with $\vec{y}=(y_{1,\bot},\cdots,y_{d-2,\bot})$.
%where $(r,\theta)$ are the polar coordinates in $(t,x)$ plane.
Before considering the slightly more complicated problem of A and B being separated by a sphere, we will warm up by computing the entanglement entropy between the region $x>0$ (subsystem A) and the region $x<0$ (subsystem B).  The entangling surface is the plane $x=0$.  We will take the system to be in its ground state.
%Suppose the action is Lorentz invariant and the interface of the two subsystems, called entangling surface, is a plane.
%The entangling surface is at $x=0$ and  are the coordinates along the entangling surface.

A convenient basis for ${\cal H}_A$ are the states $|\psi_A\rangle$, where $\psi_A=\psi_A(x)$ is a function on A.  The matrix elements of the reduced density matrix $\rho_A$ are given by the  Euclidean signature functional integral:
%
%The reduced density matrix describing the state of the field $\psi_A$ and $\psi'_A$ for the subsystem  $\mathcal{H}_{A}$
%the  has the following path integral expression
\beq
\langle \psi_A |\rho | \psi'_A\rangle ={1\over Z}\int_{\psi(t=0_-, x>0)=\psi_A\atop \psi(t=0_+, x>0)=\psi'_A} \mathcal{D}\psi 
~ e^{-S_E}%e^{-\int_{-\infty}^{+\infty}dt \mathcal{L}(\psi)},
\label{pathint-reduced}
\eeq
where $Z$ is a normalization factor.  The boundary conditions set $\psi$ to equal $\psi_A$ ($\psi_A'$) on either side of the cut at $t=0, x>0$.  %Field configurations with $x<0$ are integrated over. 
 $\psi$ is required to fall off as $t\rightarrow\pm \infty$; this puts the system in its ground state. % enforces the ground state condition.
%
%\if0
%The Hilbert space 
%which allows the path integral description.
%We choose a ground state of the system for $|\Psi_{AB}\rangle$. Then the overlap between a basis state $|\psi_A\rangle\otimes |\psi_B\rangle  \in \mathcal{H}_{A}\otimes \mathcal{H}_{B}$ and  $|\Psi_{AB}\rangle$ is
%\bea
%(\langle \psi_A|\otimes \langle\psi_B|)|\Psi_{AB}\rangle
%=
%\int \mathcal{D}\psi ~ e^{-\int_{-\infty}^{-\epsilon}dt \mathcal{L}(\psi)} \Big|_{\psi(t=-\epsilon)=\psi_A\otimes \psi_B}
%\eea
%Here, $\mathcal{L}(\psi)$ is the Lagrangian of the field $\psi$ and the boundary condition is set at time $t=-\epsilon$ where $\epsilon$ is a small positive number.
%The path integral from the past infinity to the finite time $t=-\epsilon$ in Euclidean space automatically selects the ground state $|\Psi_{AB}\rangle$.
%A similar relation is true for the conjugate
%\bea
%\langle\Psi_{AB}|(| \psi'_A\rangle |\otimes |\psi'_B\rangle)
%=
%\int \mathcal{D}\psi ~ e^{-\int_{+\epsilon}^{+\infty}dt \mathcal{L}(\psi)} \Big|_{\psi(t=+\epsilon)=\psi'_A\otimes \psi'_B}
%\eea
%By combining these two,
%the density matrix $\rho_A$ is
%\bea
%\rho_{A}(\psi_A|\psi'_A)=\int \mathcal{D}\psi \mathcal{D}\psi_B\mathcal{D}\psi'_B
%~ e^{-\int_{-\infty}^{+\infty}dt \mathcal{L}(\psi)} \Big|_{\psi(t=-\epsilon)=\psi_A\otimes \psi_B, \psi(t=+\epsilon)=\psi'_A\otimes \psi'_B}.
%\label{pathint-reduced}
%\eea
%So far the argument is general for any quantum theory.
%\fi
%

%We will consider Lorentz invariant theories.  

We can also study the system in polar coordinates
\beq
ds^2=z^2d\theta^2+dz^2+d\vec{y}^2_{\bot},
\label{flat coord}
\eeq
where $\theta$ has periodicity $2\pi$.  The boundary conditions in the path integral (\ref{pathint-reduced}) are enforced on either side of the ray $\theta=0$.  Thus we can interpret $\rho_A$ as an operator which rotates by an angle $2\pi$ in the $\theta$ direction \cite{Kabat:1994vj}
%are enforced on either side of the $\theta=0$ 
%In this case, the reduced density matrix (\ref{pathint-reduced}) can be interpreted as 
%a thermal density matrix 
\beq
\rho={1\over Z}e^{-2\pi H_E}
\eeq
where $H_E=i{\pd\over \pd \theta}$ is the Euclidean rotation operator.  If  $H_E$ is regarded as a physical Hamiltonian (an ``entanglement Hamiltonian") this is precisely a thermal density matrix at temperature $T=1/2\pi$; the normalization factor $Z$ is the usual finite temperature partition function $\Tr e^{-2 \pi H_E}$.  In Lorentzian signature $H_E$ becomes the Rindler Hamiltonian which generates Lorentz boosts in the $(x,t)$ plane, and the origin $z=0$ is the Rindler horizon associated to an accelerating observer.  %Here $H_E={\pd\over \pd \theta}$, called an entanglement hamiltonian, is a rotation generator in Euclidean space or a boost operator in Lorenzian singature. 
%Notice that this change of the interpretation is possible because the entangling surface $r=0$ is the fixed point of the boost generator. 
%Thus the "entanglement Hamiltonian" H_E is nothing more than a generator of Euclidean rotations.

Let us now specialize to conformal field theories.  This implies that, under a conformal rescaling of the metric, $\rho_A$ will change by conjugation a unitary operator; this is a trivial change of basis which we will suppress.  We can conformally rescale the metric (\ref{flat coord}) by a factor of {$z^2 / L^2$} to obtain
\beq
ds^2_{scaled}%&=&r^2d\theta+dr^2+d\vec{y}^2_{\bot}\cr
=
%{r^2\over L^2}\left(
d\tilde{\theta}^2+L^2{dz^2+d\vec{y}^2_{\bot}\over z^2}
%\right)
\label{polar coord}
\eeq
where $\tilde{\theta}=L\theta$ is periodic with period $2 \pi L$.  This metric is a  product of a circle times hyperbolic space $\mathbb{H}_{d-1}$; the size of the circle, and the size of the hyperbolic space, are set by the parameter $L$.  The reduced density matrix is
\beq
\rho_A={e^{-\tilde{H_{E}}/T_0}\over Z(T_0)}
\label{modular hamiltonian}
\eeq
where $T_0={1\over 2 \pi L}$ and ${\tilde H}_E = i{\pd \over \pd {\tilde \theta}}$ generates translations in the ${\tilde \theta}$ direction.
%$H_E=H_{\tau}/T_0+\log Z(T_0)$ and
We note that ${\tilde H}_E$ is the Hamiltonian describing time evolution for a CFT on $\mathbb{H}_{d-1} \times \mathbb{R}_t$.  $Z(T_0) =\Tr [e^{-{\tilde H}_{E}/T_{0}}]$ is the finite temperature partition of the theory in hyperbolic space.
%$Z(T_{0})$ is the thermal partition function on the hyperbolic space with the hamiltonian $H_{E}$ at the Rindler temperature $T_{0}$:
%\bea
%Z(T_{0})=\Tr [e^{-H_{E}/T_{0}}].
%\eea
%
%at temperature $T_0=1/2\pi L$, $()$. 
%The entanglement hamiltonian is now a translation generator in $\tilde{\theta}$ direction and it can interpreted as a hamiltonian on this hyperbolic space.
Note that the conformal transformation has mapped the entangling surface to the boundary of hyperbolic space. 
%The above example is a mapping from the half space on the flat space to the hyperbolic space. 

Once we consider a conformal field theory many other mappings are possible.  For example, the reduced density matrix for a CFT on a sphere, with entangling surface at the equator, can also be put in the form (\ref{modular hamiltonian}).  So can the CFT on a flat space with entangling surface equal to a sphere.  These are all related by conformal transformations
(as described in e.g. \cite{Casini:2011kv}).  More generally, any time the entangling surface is the locus of fixed points of some conformal transformation, one can interpret the entanglement Hamiltonian as the generator of that conformal transformation (in the same way that the point $z=0$ above is the fixed point of the rotation ${\pd \over \pd \theta}$).  A conformal rescaling then turns this into the Hamiltonian on hyperbolic space. 

In the following sections, we will consider the entangling surface to be a sphere of radius $L$, % such that the entanglement entropy is not IR-divergent. 
%following In this case, the conformal mapping is the 
following \cite{Casini:2011kv, Hung:2011nu}. We take the flat space-time metric to be
\beq
ds^2=-dt^2+dr^2+r^2d\Omega_{d-2}^2 \,,
\eeq
with region A being the inside of the sphere $r=L$. We now map the causal development of A to hyperbolic space times time by the following coordinate transformation:
\beq
t=L\frac{\sinh(\tau/L)}{\cosh u + \cosh(\tau/L)}, \ \ \ \ \ \ \ r=L\frac{\sinh u}{\cosh u + \cosh(\tau/L)} \,.
\eeq
The metric becomes
\beq
ds^2=\Omega^2[-d\tau^2+L^2(du^2+\sinh^2ud\Omega_{d-2}^2)]
\eeq
with $\Omega=(\cosh u + \cosh(\tau/L) )^{-1}$.  This is conformally equivalent to hyperbolic space times time, i.e. to the metric
(\ref{polar coord}). The temperature as well as the curvature of hyperbolic space are fixed by the radius of the sphere $L$. Conformal transformations act as unitary operators on the Hilbert space of the theory.  Thus the reduced density matrix is simply given by (\ref{modular hamiltonian}), except conjugated by some Unitary operator $U$:
\beq
\rho_A=U {e^{-\tilde{H_{E}}/T_0}\over Z(T_0)} U^{-1} .
\label{modular hamiltonian 2}
\eeq

%
%Conformal transformations of this sort allow other mallow other mappings to the hyperbolic space: a northern/southern hemisphere on $S^{d}$, and
%a spherical region on a flat space. The radius of the hyperbolic space $L$, is related
%to the radius of the sphere $S^{d}$ and the size of the spherical region in each mapping. In any case, the entangling surface is the fixed point of the Lorentz and the conformal transformations.
%Conformal transformation also allows us to choose a spherical region, instead of the half space, as a subsystem $A$ \cite{Casini:2011kv}. 
%Starting with a flat space metric
%\bea
%ds^2=-dt^2+dr^2+r^2d\Omega^2_{d-2},
%\eea
%the coordinate transformation
%\bea
%t&=&L{\sinh(\tau/L)\over \cosh u+\cosh(\tau/L)},\cr
%r&=&L{\sinh u\over \cosh u+\cosh(\tau/L)},
%\eea
%brings us to
%\bea
%ds^2={1\over (\cosh u+\cosh(\tau/L))^2}(-d\tau^2+L^2(du^2+\sinh^2 ud\Omega^2_{d-2}))
%\eea
%Notice that the sphere $r=L$, $t=0$ is the fixed point of Lorentz + conformal transformation. Therefore, the same argument as above applies and we can choose it as the entangling surface.

We can now use the equivalence between the reduced density matrix and the thermal density matrix on the hyperbolic space to relate the entanglement entropy and R\'{e}nyi entropy to the thermal entropy in hyperbolic space.
The entanglement entropy $S_{EE}$ and the R\'{e}nyi entropy $S_{n}$  are defined as
\beq
S_{EE}=-{\Tr}_A \rho \log \rho,
\eeq
and 
\beq
S_{n}={1\over 1-n}\log \Tr [\rho^{n}].
\label{Rnyi tr}
\eeq
We call $n$ the R\'{e}nyi parameter.  If $S_n$ is analytic near $n=1$, then $S_{EE} = \lim_{n\to1} S_{n}$.
%$\al$, the entanglement entropy is 
%obtained by taking $\al\to1$ limit.
%From the positivity and the hermiticity of the reduced density matrix, we can introduce an hermitian operator $H_{E}$, called
% entanglement hamiltonian
%\bea
%\rho = e^{-H_{E}}.
%%\label{modular hamiltonian}
%\eea
%As explained above, the reduced density matrix is essentially equivalent to the thermal density matrix on the hyperbolic space.
%To make the analogy to the standard thermodynamics clearer, we write it in the following form 
%\bea
%\rho={e^{-f{E}/T_0}\over Z(T_0)}
%\label{modular hamiltonian}
%\eea
%where 
%%$H_E=H_{\tau}/T_0+\log Z(T_0)$ and
%$Z(T_{0})$ is the thermal partition function on the hyperbolic space with the hamiltonian $H_{E}$ at the Rindler temperature $T_{0}$:
%\bea
%Z(T_{0})=\Tr [e^{-H_{E}/T_{0}}].
%\eea
The trace of the $n$-th power of $\rho$ is then given by thermal partition function on a hyperbolic space with radius $L$
at temperature $T=1/2\pi Ln$:
\beq
\Tr[\rho^{n} ]={\Tr [e^{-n H_{E}/T_{0}}]\over Z(T_0)^{n}}={Z(T_0/n)\over Z(T_0)^{n}}.
\label{renyi-partition}
\eeq
Thus the R{\'e}nyi entropy is
\beq
S_{n}={n\over 1-n}{1\over T_{0}}[F(T_{0})-F(T_{0}/n)].
\label{Rnyi-therm-free}
\eeq
where $F(T)=-T\log Z(T)$ is the free energy of a CFT on $\mathbb{H}_{d-1}$.  
This can be written as
\beq
S_{n}={n\over n-1}{1\over T_{0}}\int_{T_{0}/{n}}^{T_{0}}S_{\text{therm}}(T)dT.
\label{Rnyi-therm}
\eeq
where 
\beq
S_{\text{therm}}(T)=-\pd F(T)/ \pd T
\label{thermal S}
\eeq 
is the usual thermal entropy on $\mathbb{H}_{d-1}$.

We are interested in computing the entanglement spectrum (i.e. the eigenvalue spectrum of the reduced density matrix).
In general this will have both a discrete and continuous part.
In the basis that diagonalizes the reduced density matrix, the R{\'e}nyi entropy is
\beq
S_{n}%&=&{1\over 1-\al}\log \Tr [\rho_A^{\al}]\cr
={1\over 1-n}\log  \left[ \sum_{i} \bar{d}_{i}\lambda_{i}^{n} + \int_{0}^{1} d\lambda d(\lambda) \lambda^{n}  \right]
\label{discrete Rnyi}
\eeq
where $\lambda_{i}$ are the eigenvalues of $\rho_A$ with $\lambda_1>\lambda_{2}>\cdots$, and $\bar{d}_{i}$  ($d(\lambda)$)  are the discrete (continuous) degeneracies of the eigenvalues $\lambda_{i}$ and $\lambda$.
Note that the eigenvalues of the entanglement hamiltonian $H_E$, which we denote $h_{i}=-\log{\lambda_{i}}$, satisfy $h_{1}<h_{2}<\cdots$.
Since $\lambda_i$ and $\lambda$ are equal or smaller than 1, the R{\'e}nyi entropy is a monotonically decreasing function of $n$. Thus, if we expand $S_n$ in powers of $n$, it will contain only constant terms or negative powers of $n$. In the large $n$ limit, only the constant term survives; it gives the largest eigenvalue $\lambda_1$:
%
%\bea
%S_{\al}=\text{const}-\sum_{n}{1\over \al^n}S^{n}
%\eea
%
\beq
S_{\infty}=-\log\lambda_{1}.
\label{min-ent}
\eeq

%
%Since the thermal entropy is also defined by the same function, replacing the reduced density matrix by the thermal density matrix, the above argument
%suggests that
%the entanglement entropy $S_{EE}$ of the original theory is mapped to the thermal entropy $S_{\text{therm}}$ on  $S^1\times H_{d-1}$ at the temperature $T=T_0=1/2\pi L$. In this paper, we call the above method of computing the reduced density matrix as ``the Rindler method."
%In general, this method should be distinguished from the replica method. The replica method does not require the Lorentz invariance and can be used for 
%arbitrary entangling surface, while $\alpha$ has to be restricted to be an integer.   
%On the other hand, the Rindler method is restricted to a planar or conformally planar entangling surface in relativistic or conformal field theory. In this method, $\al$
%can take any real number.

%\subsection{UV and IR divergences}

We note that the entanglement entropy in a continuum theory is UV divergent unless one employs a cutoff near the entangling surface.
Likewise, the thermal entropy on hyperbolic space diverges because of the infinite volume unless one employs an IR cutoff.\footnote{In general, the theory may also have standard UV divergences which have nothing to do with the entangling surface.  We assume that these are regularized in ordinary ways, e.g. via  zeta function regularization or background subtraction.}
One can show that these two divergences are essentially the same thing; they are mapped to one another by the conformal transformation \cite{Casini:2011kv}. 
\section{Holographic R{\'e}nyi entropies}
\label{HEE and HRE}

We now consider the computation of R\'{e}nyi entropies in CFTs with bulk gravity duals, where they are related to entropies of hyperbolic black holes.  We will describe the instabilities of these black holes in section 3.2, using a simple near-horizon analysis of the extremal black hole.

\subsection{Hyperbolic Black Hole}
%Based on the equivalence between the reduced density matrix and the thermal density matrix discussed above, Casini, Huerta and Myers \cite{Casini:2011kv} proposed a partial derivation of the holographic computation of entanglement entropy. 
%
%We have seen that 
%In the case of conformal field theory, the entanglement entropy of a spherical region with radius $L$ is 
%equivalent to the thermal entropy on $ S^1\times H_{d-1}$ with $S^1$ periodicity be $2\pi L$ and the radius of $H_{d-1}$ be $L$.
%The gravity dual of the thermal system on the hyperbolic space is an AdS black hole space-time with a hyperbolic horizon
%\bea
%ds^2=-\left({r^2\over L^2}-1  \right)d\theta^2+{dr^2\over {r^2\over L^2}-1}+r^2dH^2_{d-1}.
%\label{Rinder-AdS}
%\eea
%where $L$ is the AdS radius.
%This space-time is in fact a hyperbolic slicing of AdS space and the horizon at $r=L$ is the Rindler horizon.
%It covers half of the global AdS space.
%The entanglement entropy and the thermal entropy are given by Bekenstein-Hawking formula
%\bea
%S_{EE}=S_{thermal}={r_{H}^{d-1}\text{vol}(H_{d-1})\over  4G_{N}}.
%\eea
%The area of the hyperbolic horizon is infinity and we need an IR cutoff near the boundary of the hyperbolic space.
%This IR cutoff should be identified with the IR cutoff of the thermal entropy on the hyperbolic space, and the UV cutoff of the entanglement entropy
%at the entangling surface \cite{Casini:2011kv}.

In a CFT, the  R{\'e}nyi entropies $S_n$ are related to the thermal entropies $S(T)$ of the theory on hyperbolic space $\mathbb{H}_{d-1}$ times time.  In particular, from equation (\ref{Rnyi-therm}}), $S_n$ is an integral of $S(T)$ from $T=T_0=1/2\pi L$ to $T=T_0/n$.  Thus we must compute $S(T)$ for $T$ between $0$ and $T_0$.

We will consider CFTs with gravity duals.  
In this case the thermal entropy $S(T)$ is equal to the entropy of a black hole in $AdS_{d+1}$
with hyperbolic horizon and Hawking temperature $T$  \cite{Casini:2011kv, Hung:2011nu}:
\beq
S_{therm}={r_{H}^{d-1}\text{vol}(\H_{d-1})\over  4G_{N}}
\eeq
Here $r_H$ is the value of the radial coordinate at the event horizon.
It is straightforward to construct such solutions in Einstein gravity with a negative cosmological constant.  The metric is
\beq
ds^2=-f(r)d\theta^2+{dr^2\over f(r)}+r^2dH^2_{d-1}.
\label{finite-T top bh}
\eeq
where
\beq
f(r)={r^2 \over L^2} -1-{m\over r^{d-2}},
\eeq
and $m$ is the black hole mass. The AdS metric (in Rindler-like coordinates) is recovered by setting $m=0$.
The horizon location $r_{H}$ is determined by 
\beq
f(r_H)=0.
\eeq
The black hole temperature is 
\beq
T={f'(r_H)\over 4\pi}={1\over 2\pi L}\left({r_{H}\over L}+{d-2\over 2}{mL\over  r_{H}^{d-2}}\right) \label{TBH}
\eeq

Asymptotically, the black hole metric becomes
\beq
ds^2\sim \left(r^2 \over L^2\right)\left(-d\theta^2+L^2 dH_{d-1}^2\right)~~~~~{\rm as}~ r\to\infty
\eeq
Thus the black hole is dual to the CFT on $\R\times \H_{d-1}$ at temperature $T$ given by (\ref{TBH}).  The black hole entropy therefore computes the thermal entropy of the CFT, which is in turn related to the R\'enyi entropy via (\ref{Rnyi-therm}).
Note that the AdS radius $L$ has been related to the radius of the entangling surface.

The relation between the R\'{e}nyi parameter $n$ and the mass $m$ is
\beq
n=\left({r_{H}\over L}+{d-2\over 2}{mL\over r_{H}^{d-2}}\right)^{-1}
\eeq
For the R\'{e}nyi parameter $n>1$, the black hole mass $m$ needs to take a negative value.
The extremal case is 
\beq
r^{\text{ext}}_{H}=\sqrt{{d-2\over d}}L<L
\eeq
at which the black hole temperature becomes zero. 
The black hole exists only for
\beq
r_{H}\geq r^{\text{ext}}_{H}.
\eeq
In the extremal limit, the near horizon geometry becomes $AdS_{2}\times \H_{d-1}$ with $AdS_2$ radius $L/\sqrt{d}$, and the temperature $T\to0$.

\subsection{Phase transitions: Analytic Estimates}
\label{Phase-Trans}

So far we have considered only vacuum solutions of Einstein gravity.  Let us consider the case where the CFT contains a scalar operator of dimension $\Delta$, which in the bulk corresponds to a scalar field of mass $\mu^2L^2=\Delta(\Delta-d)$. 
If $\Delta$ is sufficiently small, then it is possible for  asymptotically AdS black holes to be unstable at low temperature \cite{Dias:2010ma, Gubser:2008zu}.  This would lead to a phase transition where the scalar field condenses.

In the case of the hyperbolic black holes described above, a similar instability was considered in \cite{Dias:2010ma}.  The authors of \cite{Dias:2010ma} considered only modes which are constant on the hyperboloid, which are normalizable only if the hyperboloid is quotiented to form a compact space.  In the present case the hyperboloid is not quotiented.  We will therefore consider here a more general class of instabilities, where a normalizable mode on the hyperboloid condenses.

%Let us first consider the scalar to be constant along the hyperboloid. This was done in \cite{Dias:2010ma} where the hyperboloid was taken to be compact. In our set up, the hyperbolic slices are non-compact and this implies that the scalar is \textit{not} normalizable and corresponds to boundary conditions where we allow the dual operator to be sourced at the entangling surface.

The physics of the problem is easy to understand.
Consider a minimally coupled scalar field $\Phi$ with mass $\mu$, which behaves asymptotically as 
%The asymptotic behaviour of the scalar field is
\beq
\Phi \sim {a_{1}\over r^{\Delta_{+}}}+{a_{2}\over r^{\Delta_{-}}}+\cdots \label{asymptotics},~~~~~
\Delta_{\pm}={d\over2}\pm\sqrt{{d^2\over 4}+\mu^2L^2}.
\eeq
%The coefficients $a_1$ and $a_2$ correspond to the expectation values of conformal dimension $\Delta_{+}$ and $\Delta_{-}$ operators.
%For sufficiently high temperature black holes (in particular, for those with $T>T_0$) the Einstein black hole is always the 
We will consider cases where the mass $\mu$ is between the Breitenlohner-Freedman (BF) bound of $AdS_{d+1}$ and $AdS_{2}$,
\beq
-{d^2\over 4L^2}\leq \mu^2 <-{d\over 4L^2}.
\eeq

Let us first consider the case of a mode which is constant on the hyperboloid.
In this case, for sufficiently high temperature black holes -- in particular when $T>T_0$ -- the scalar field will always be stable in the black hole background.\footnote{This is easiest to see by noting that the $T=T_0$ black hole is just AdS${}_{d+1}$, which is stable, and that black holes only become more stable as the temperature is increased. }  Thus the dominant saddle point will be the one with $a_1=a_2=0$.  However, as $T \to 0$ the near horizon geometry becomes $AdS_2$ and the scalar field is unstable.  Thus there is a critical temperature $T_c < T_0$ at which $\Phi$ becomes unstable.  Below $T_c$, either  $a_1$ or $a_2$ (depending on which boundary conditions we choose for the scalar field) will become non-zero.  The dominant solution is a hairy black hole, which must be obtained numerically.

This instability was discussed in  \cite{Dias:2010ma}, who obtained the phase diagram  as a function of the temperature and the scalar's mass.  As expected, for $\mu$ at the $AdS_2$ BF bound, the extremal ($T\to 0$) black hole is unstable.  As $\mu^2$ is decreased the critical temperature $T_c$ increases until we reach the $AdS_{d+1}$ BF bound, which corresponds to the threshold of instability for the massless black hole (which has $T=T_0$).  We note that, as long as the field obeys the $AdS_{d+1}$ BF bound, the locally AdS solution with $T=T_0$ will be stable. We will reproduce these results in the next section, where they will be summarized in Fig. \ref{Tc_vs_Delta}. This condensation will generate a phase transition in the thermal entropy and from (\ref{Rnyi-therm}), a phase transition in the R\'{e}nyi entropy.\footnote{These results are for standard (Dirichlet) boundary conditions for the scalar field ($a_2=0$).  However, it is also possible to consider non-standard (Neumann, $a_1=0$) boundary conditions.  In this case the critical temperature continues to increase as $\Delta_-$ decreases. In fact, the critical temperature is {\it higher} than that of the massless black hole. The massless black hole is the one which computes the entanglement entropy. Thus the entanglement entropy would no longer given by the area of the hairless black hole!  Note, however, that this result is true only if we include the (non-normalizable) constant mode.  When we restrict to normalizable modes below, the massless black hole will remain hairless.  This emphasizes the important qualitative differences between the constant and normalizable modes. Related considerations are discussed in \cite{topological-instability}.}

The above analysis, however, is incomplete because it considers a mode which is non-normalizable on the hyperboloid.  
%
%As we mentionned in section \ref{boundaryconditions}, there are different boundary conditions one may impose at the entangling surface when calculating the reduced density matrix. In the previous subsection, we have allowed the possibility of an operator being sourced at the entangling surface. From the bulk point of view, this corresponded to the scalar mode that is constant on the hyperboloid, and precisely does not vanish at the boundary, i.e. the entangling surface.
%
Such modes are easy to study, as they preserve the symmetries of the hyperboloid.  Only normalizable modes, however, will lead to true instabilities of the black hole.  On a hyperboloid $\H_{d-1}$, normalizable modes can be expanded in eigenfunctions of the hyperbolic laplacian, $\nabla^2_{\H_{d-1}} \phi=-\lambda\phi$ with $\lambda>(d-2)^2/4$. 
%Let us analyse the near horizon physics more precisely. The metric of the black hole (in dimensionless units this time) is
%\beq
%\frac{ds^2}{\ell^2}=-f(r)dt^2+\frac{dr^2}{f(r)}+r^2dH_{d-1}^2
%\eeq
%with
%\beq
%f(r)=-1+r^2-\frac{r_+^{d-2}(-1+r_+^2)}{r^{d-2}}
%\eeq
%where $r_+$ is the location of the horizon (we have chosen here to express the mass of the black hole as a function of the location of the horizon). The extremal black hole with vanishing temperature is given by
%\beq
%r_+=\sqrt{\frac{d-2}{d}}
%\eeq
The extremal black hole has a near horizon region $AdS_2\times \H_{d-1}$ with radii
\beq
L_{AdS_2}=\frac{L_{AdS_{d+1}}}{\sqrt{d}} \ \ \ \ \ \ \ \ \ L_{\H_{d-1}}=\sqrt{\frac{d-2}{d}}L_{AdS_{d+1}}
\eeq
%This near horizon argument follows that of \cite{Dias:2010ma}. It appears thus natural to take the effect of the non-constant mode to be simply a shift in the mass of the scalar\footnote{This was also noted in\cite{Dias:2010ma} for rotating black holes and non-constant scalars on the transverse directions}. 
The full Laplacian is the sum of the $AdS{}_2$ and $\H_{d-1}$ Laplacians.  Thus, for a mode which is an eigenfunction of  $\nabla^2_{\H_{d-1}}$, the effective mass of the field in $AdS{}_2$ is shifted.
We therefore expect an instability when
\beq
\mu^2+\frac{\lambda}{L^2_{\H_{d-1}}}\leq-\frac{1}{4L^2_{AdS_2}}
\eeq
where $\lambda$ is the lowest eigenvalue of the laplacian on $\mathbb{H}_{d-1}$.  
This occurs when the scalar masses are in the following range:
%We also want the scalar field to obey the $AdS_{d+1}$ BF bound so we get the following range of masses:
\beq
-\frac{d^2}{4}\leq \mu^2L^2_{AdS_{d+1}}\leq-\frac{d(d-1)}{4} \label{massbound}
\eeq
There will be masses in this range for any dimension $d$. For example in $AdS_5$ we will find an instability when
\beq
-4\leq \mu^2L^2_{AdS_{5}}\leq-3
\eeq
We will now turn to a numerical analysis of this instability.

%\section{Spectral function}
%\label{Spectral Func}

\section{The Constant Mode and a Hairy Black Hole}

We will now study numerically the instability of the hyperbolic black hole to the development of scalar hair.
We will begin with a discussion of the constant mode on the hyperboloid.  This mode is particularly simple, as it preserves all of the hyperbolic symmetries.  Thus the hairy black hole can be constructed relatively easily using a full non-linear numerical method.  However, this mode is non-normalizable so will not fluctuate in the gravity theory.  The results of this section can therefore be viewed as a simplified model (one where the hairy black hole can be constructed explicitly) of the more complicated case considered in the next section.  

We note that, although the constant mode considered in this section is non-normalizable on the hyperboloid $\H_{d-1}$, it is normalizable on compact quotients of $\H_{d-1}$.  Moreover, one may wish to consider entanglement entropies with insertions of operators at the entangling surface (the boundary of the hyperboloid) which source the scalar field.  If these operators turn on the constant mode of the scalar, then the results of this section would be precisely correct, rather than just a simplified model.   

To present numerical results we focus on the specific case of  a scalar field theory coupled to Einstein gravity in 5 dimensions.\footnote{We will consider charged  black holes in \cite{charged-hyperbolic,Belin:2014mva}.}  Our numerical approach follows \cite{Dias:2010ma} closely.
The action is
\beq
S={1\over 16\pi G_{N}}\int d^5x\sqrt{-g}
\left[
R+{12\over L^2}-({\nabla \Phi})^2-\mu^2 \Phi^2
\right]
\eeq
where $\mu$ is a mass of the scalar field.
We use the following metric  and scalar field ansatz 
\beq
ds^2=-f(r)e^{2\chi(r)}dt^2+{dr^2\over f(r)}+r^2dH^2_{3},~~~~~\Phi=\Phi(r). \label{ansatz}
\eeq
We are considering only constant modes on the hyperboloid; such modes will not fall off at the boundary of $\mathbb{H}_{d-1}$, i.e. the entangling surface.  
%Thus they describe contributions to the reduced hamiltonian where the scalar field has a non-trivial vev at the entangling surface.%\footnote{It is also interesting to look for non-constant modes on the $\mathbb{H}_{d-1}$; we will leave this for future work.}

The equations of motion are
\bea
&&\Phi''(r)+{1\over  r f(r)}
\left[
\left(-{\mu^2\over3}r^2\Phi(r)^2+f(r)+{4r^2\over L^2}
-2\right)\Phi'(r)
-\mu^2r\Phi(r)
\right]=0, \label{eomf}\\
&&f'(r)+
\left(
{2\over r}+{r\over 3}\Phi'(r)^2
\right)f(r)+{\mu^2\over 3}r\Phi(r)^2+{2\over r}-{4r\over L^2}=0, \label{eomphi}\\
&&\chi'(r)-{r\over 3}\Phi'(r)^2=0 \label{eomchi}.
\eea
For an asymptotically AdS$_5$ space-time, the scalar field behaves as (\ref{asymptotics}). For $a_1=0$, $a_2$ gives the expectation value of conformal dimension $\Delta_{-}$ operator, and for $a_2=0$, $a_1$ gives the expectation value of 
conformal dimension $\Delta_{+}$ operator. In this section, we set $a_2=0$; this is the ``standard" (i.e. Dirichlet) boundary condition, where the mode which falls of more slowly at $r\to\infty$ is set to zero.
At the AdS$_5$ BF bound, i.e., $\mu^2L^2=-4 ~(\Delta=2)$, $\Delta_{\pm}$ degenerate and the asymptotic behaviour of the scalar field becomes
\beq
\Phi(r) \sim {a_1\over r^{2}}+{a_2 \log (r)\over r^{2}}+\cdots
\eeq
In this case, we set $a_{2}=0$.

The metric functions behave as
\beq
f(r\to \infty)={r^2\over L^2}-1+{r_0^2\over r^2},~~~\chi(r\to \infty)=\mathcal{O}(r^{-2\Delta_{+}}),
\eeq
%To solve these equations, we need to set boundary conditions.
%For the simplicity of the numerical computation, we set the horizon location $r_{+}$, the metric and the scalar functions at the horizon
%$f(r_{+})=0$ and $\Phi(r_{+})=a$, where $a$ is a constant.
%The value of $\Phi'(r_{+})$ is determined from (\ref{eomf}).
%We choose the mass to be BF bound, i.e., $\mu^2L^2=-4 ~(\Delta=2)$.
%
%\if0
%In this case, the solutions must behave asymptotically as
%\bea
%f(r\to \infty)={r^2\over L^2}-1+{r_0^2\over r^2}, ~~~ \Phi(r\to\infty)={\phi_0L^2\over r^2},~~~\chi(r\to \infty)=\mathcal{O}(r^{-4}).
%\eea
%\fi
%
and the thermodynamical quantities are
\beq
T={f'(r_{H})e^{\chi(r_{H})}\over 4\pi},~~~S_{\text{therm}}={r_{H}^3\text{vol}(\H_{3})\over 4G_N},~~~E={(-3r_{0}^2+\delta_\Delta^22(a_1^2/L^2))\text{vol}(\H_{3})\over 16\pi}
\label{thero-dy quan}
\eeq

We have numerically investigated hairy black hole solutions.  For each value of $\mu^2 L^2 <-1$ we find hairy black holes below a critical temperature $T_c$.  When they exist, the hairy black holes have lower free energy than the Einstein black hole.  Thus the scalar condensate phase ($a_1\neq 0$) is thermodynamically favoured.  
When the scalar mass is equal or greater than the AdS$_2$ BF bound ($\mu^2L^2\ge -1$, $\Delta \ge 2+\sqrt{3}$), the Einstein black hole is stable and the expectation value of the scalar field ($a_1$) is always zero.

\begin{figure}[htb]
\begin{center}
\includegraphics[width=.45\textwidth]{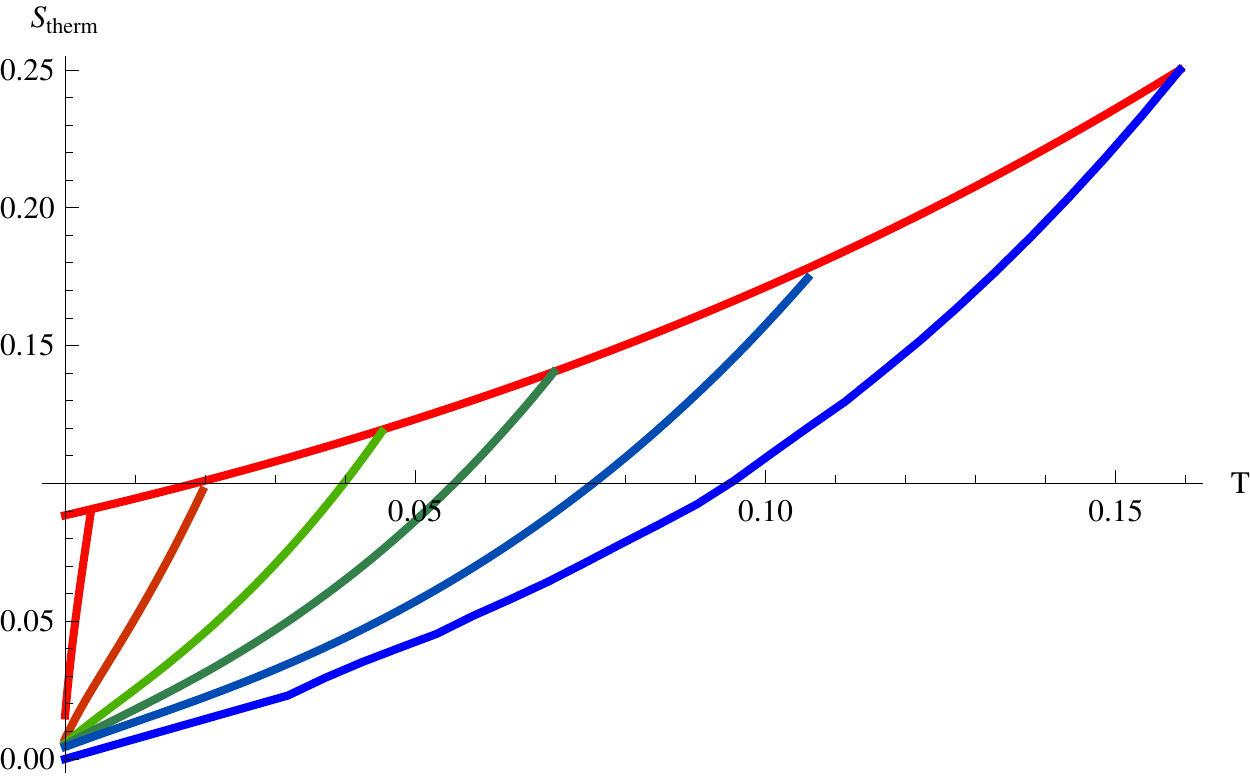}
\caption{Thermal entropy $S_{therm}$ as a function of the temperature $T$.
The upper (red) curve is for the unbroken phase (Einstein black hole).
%When the mass is equal or greater than the AdS$_2$ BF bound ($\mu^2L^2\ge -1$, $\Delta \ge 2+\sqrt{3}$), the state is always in this unbroken phase.
The lower (light red, orange, light green, green, blue-green and blue) curves are  for hairy black holes with
$\mu^2L^2= -2.2, -3,-3.5,-3.75, -3.9375, -4$ ($\Delta=3.34,3,2+1/\sqrt{2},2.5, 2.25, 2$). 
The critical temperatures are $T_c=0.0037,~0.020, ~0.045,~0.070, 0.106, 1/2\pi$.
Note the lower right (blue) curve is for the scalar at the AdS$_5$ BF bound, for which the critical temperature is that of the massless black hole.
}
\label{temperaentro-all}
\end{center}
\end{figure}

\begin{figure}[htb]
  		 \begin{minipage}[b]{0.45\linewidth}
     			\centering \includegraphics[scale=0.9]{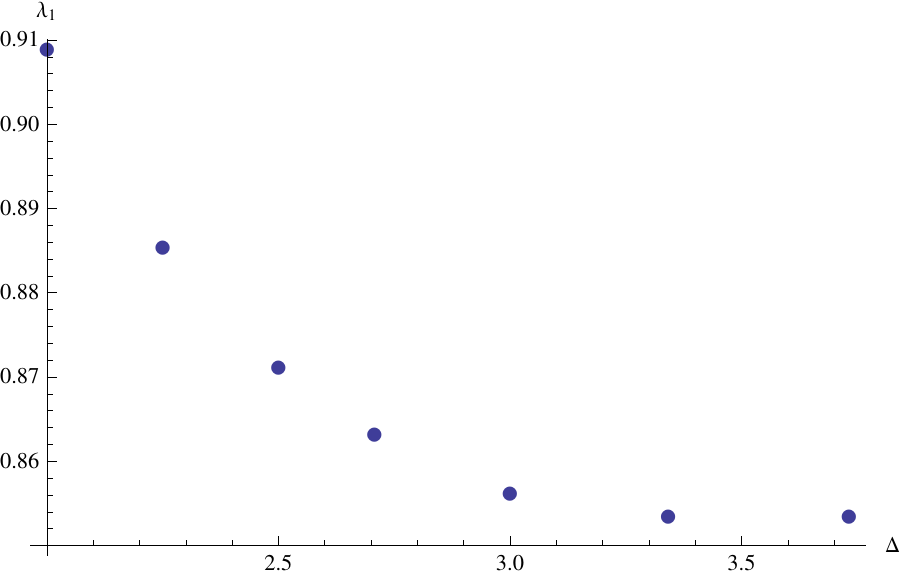}
      			\caption{Lowest eigenvalue of $\rho$ as a function of conformal dimension.}
			\label{dim-lambda1}
   		\end{minipage}\hfill
   		\begin{minipage}[b]{0.45\linewidth}   
     			 \centering \includegraphics[scale=0.9]{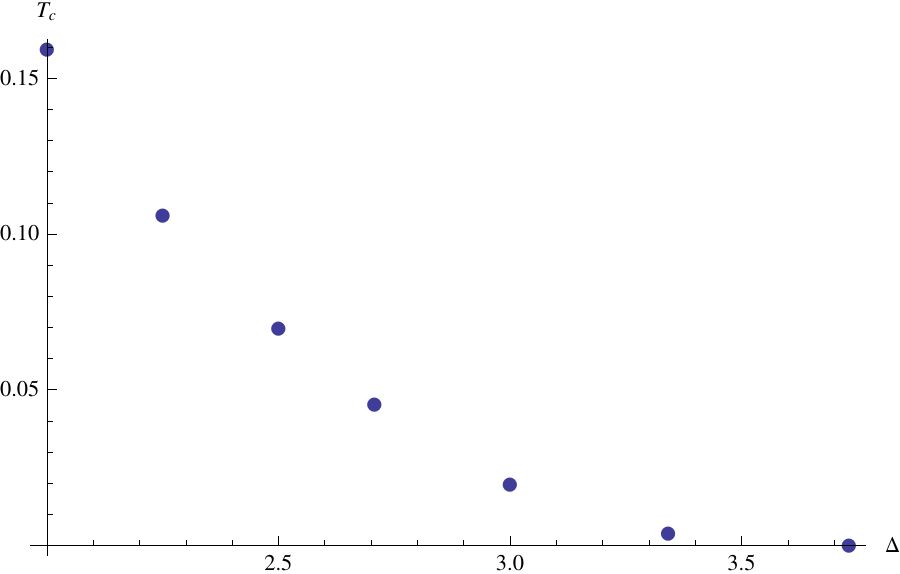}
     			 \caption{ Critical temperature as a function of conformal dimension.}
			\label{Tc_vs_Delta}
 		  \end{minipage}
	\end{figure}

Our results are presented in Fig \ref{temperaentro-all}, which shows the thermal entropy as a function of the temperature.\footnote{In this graph we set $L=\rm{vol}(\H_3)=1$ (i.e. we plot the entropy density).} 
%The area law and the information of the cutoff and the central charge dependence can be recovered by keeping track of the $L$ and  vol$(H_3)$ dependence explicitly. 
 The Einstein black hole is the top curve (red) in Fig. \ref{temperaentro-all}.  We have also plotted in Fig. \ref{temperaentro-all} the entropy of the hairy black hole for a variety of masses below the AdS${}_2$ BF bound.
 The critical temperature $T_c$ at which the scalar condenses is where these curves meet the Einstein black hole curve.
 This temperature gets larger as we decreases the scalar mass.  At the AdS$_5$ BF bound ($\mu^2L^2= -4$, $\Delta = 2$), the critical temperature becomes the Rindler temperature $T_c=T_0={1\over 2\pi L}$.  This is the blue line in Fig. \ref{temperaentro-all}.
 Note that the hairy black hole has smaller thermal  entropy, but lower free energy, than the Einstein black hole.
% This is not a problem since it is the free energy, rather than the thermal entropy, that we need to compare.
% Also it is consistent with Ryu-Takayanagi formula which states that if there are more than one extremal surfaces ending on the entangling surface,
% we need to choose the one with the smallest area.
Fig. \ref{Tc_vs_Delta} shows the critical temperatures as a function of $\Delta$, the dimension of the lowest dimension scalar operator.

\begin{figure}[htb]
\begin{center}
\includegraphics[width=.45\textwidth]{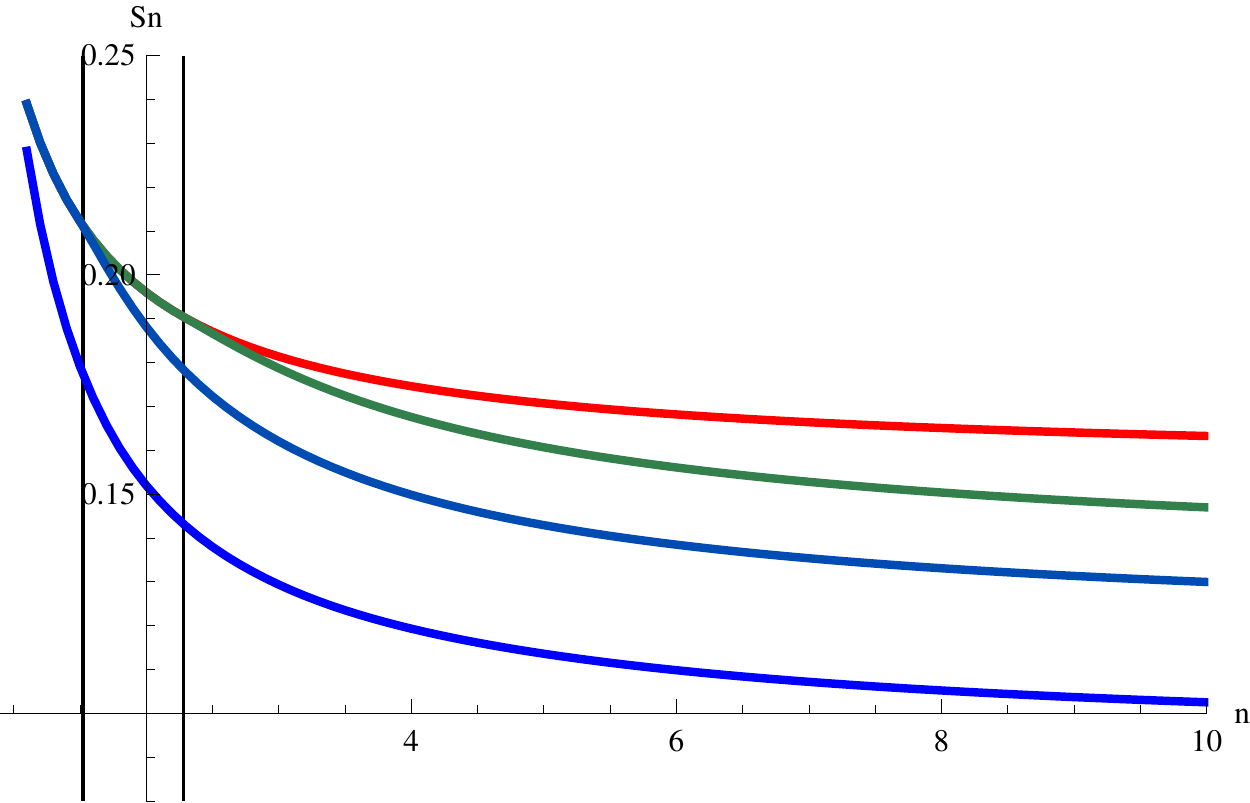}
\caption{R{\'e}nyi entropy as a function of $n$ for $\mu^2L^2=-1,-4,-3.75$ and $-3.9375$.
The phase transition is at $n_{crit}\simeq 2.28$ for $\mu^2L^2=-3.75$  and $1.52$ for
$\mu^2L^2=-3.9375$.
}
\label{renyi-com}
\end{center}
\end{figure}

From these thermal entropies we can compute the R\'{e}nyi entropy, via  (\ref{Rnyi-therm})
\beq 
S_{n}={n\over n-1}{1\over T_{0}}
\left(
\int_{T_{0}/{n}}^{T_{crit}}S^{ES}_{\text{therm}}(T)dT
+
\int_{T_{crit}}^{T_{0}}S^{E}_{\text{therm}}(T)dT
\right),
\label{Rnyi-therm-2}
\eeq
where $S^{ES}_{\text{therm}}(T)$ is the entropy of the hairy black hole (the broken phase) and 
$S^{E}_{\text{therm}}(T)$ is the entropy of the Einstein black hole (the unbroken phase).
The R{\'e}nyi entropy as a function of $n$ is plotted in Fig. \ref{renyi-com}. 
As the derivative of the thermal entropy with respect to the temperature is discontinuous, the second derivative with respect to $n$ of the R{\'e}nyi entropy is discontinuous (we have integrated once)\footnote{Just as in the holographic superconductor, the condensation of the scalar field is a second order phase transition \cite{Gubser:2008zu}. This translates into the discontinuity of the second derivative of $S_n$ with respect to $n$.}.
The upper (red) curve is the Einstein black hole.     The other curves describe cases where a scalar field condenses at some value of $n$.  Note that the R\'{e}nyi entropy $S_n$ will always approach the Einstein black hole result as $n\to 1$, because the massless black hole is stable for any scalar obeying the BF bound.  
%a hairy black hole leads to 
%are masses green line is for 
%$\mu^2L^2=-3.75$. The R{\'e}nyi parameter corresponding to the critical temperature $\alpha_{crit}=T_0/T_c$ is $2.28$.
%As can be seen, the R{\'e}nyi entropy coincides with that of the Einstein for $\al<\al_{crit}$ and deviates toward the Einstein-scalar
%above $\al_{crit}$.
As one decreases the mass, the critical temperature gets higher and the asymptotic value of the R{\'e}nyi entropy at large $n$ gets farther from the Einstein result.
%gets closer to that of the Einstein-scalar.

%%%%% begin Figure
%%%%% end Figure

%%%%% begin Figure

The spectral function $\bar{d}_i$ and $d(\lambda)$ in (\ref{discrete Rnyi}) can be obtained from the inverse Laplace transformation
%\footnote{
%Laplace transformation is defined by
%\bea
%F(s)=\int_{0}^{\infty}f(t)e^{-ts}dt.
%\eea
%This transformation is well-defined as long as the function $f(t)$ is piecewise continuous.
%The spectral function $d(\lambda)$ is discontinuous and takes zero for $\lambda\le \lambda_1$, but still it satisfies this condition.
%So we assume that $d(\lambda)$ is defined in $\lambda\in [0,1]$. In this case, by changing the variables $e^{-t}=\lambda \in [0,1]$,
%we rewrite the Laplace transformation as
%\bea
%F(s)=\int_{0}^{1}\left({f(-\log(\lambda))\over \lambda}\right)\lambda^{s}d\lambda.
%\eea
%So the spectral function is
%\bea
%d(\lambda)={f(-\log(\lambda))\over \lambda}.
%\eea
%} 
\beq
\exp\left((1-n)S_{n}\right)=\sum_{i} \bar{d}_{i}\lambda_{i}^{n} + \int_{0}^{\lambda_1} d\lambda d(\lambda) \lambda^{n}. 
\label{reny-spectral}
\eeq
We can now use our numerical results to plot the spectral density $d(\lambda)$ (including the discrete part $\bar{d}_i$ in $d(\lambda)$ by allowing a delta function).  
We will focus on the two simplest cases: $\mu^2L^2= -1$ and $\mu^2L^2= -4$, for which the critical temperatures are $T_c=0$ and $T_c=T_0=1/2\pi L$.
%From the regularity of the R\'{e}nyi entropy in large $\al$ limit,
%

We first note that, since Renyi entropies are UV divergent, the eigenvalues are exponentially suppressed with the cutoff.  For example, as explained in \cite{Hung:2011nu}, the eigenvalue $\lambda_1$ scales as
\beq
\lambda_1\sim e^{-\text{const.}\times C_T\times\mathcal{A}/\epsilon^{d-2}}
\eeq
where $\mathcal{A}$ and $\epsilon$ are respectively the area of the entangling surface and the UV cutoff.  Here $C_T$ is related to the 2-point function of the stress tensor and counts the degrees of freedom of the theory, which is typically taken to be large for gravity to be classical in the bulk. For even dimensional CFTs, it is related to the A-type trace anomaly. 
%We are interested in computing the constant in front of these large factors which captures information about the particular CFT we are considering, in our case wether a condensation arises or not. 
%In terms of the gravity calculation,
In the bulk gravity calculation, $C_T$ is set by the AdS radius in Planck units ($C_T\sim(L/\ell_p)^{d-1}$)
and the UV divergence becomes the volume divergence of the hyperboloid.  We may therefore introduce a UV regulator by taking the volume of the hyperboloid to be finite.  

\begin{figure}[htb]
\begin{center}
\includegraphics[width=.45\textwidth]{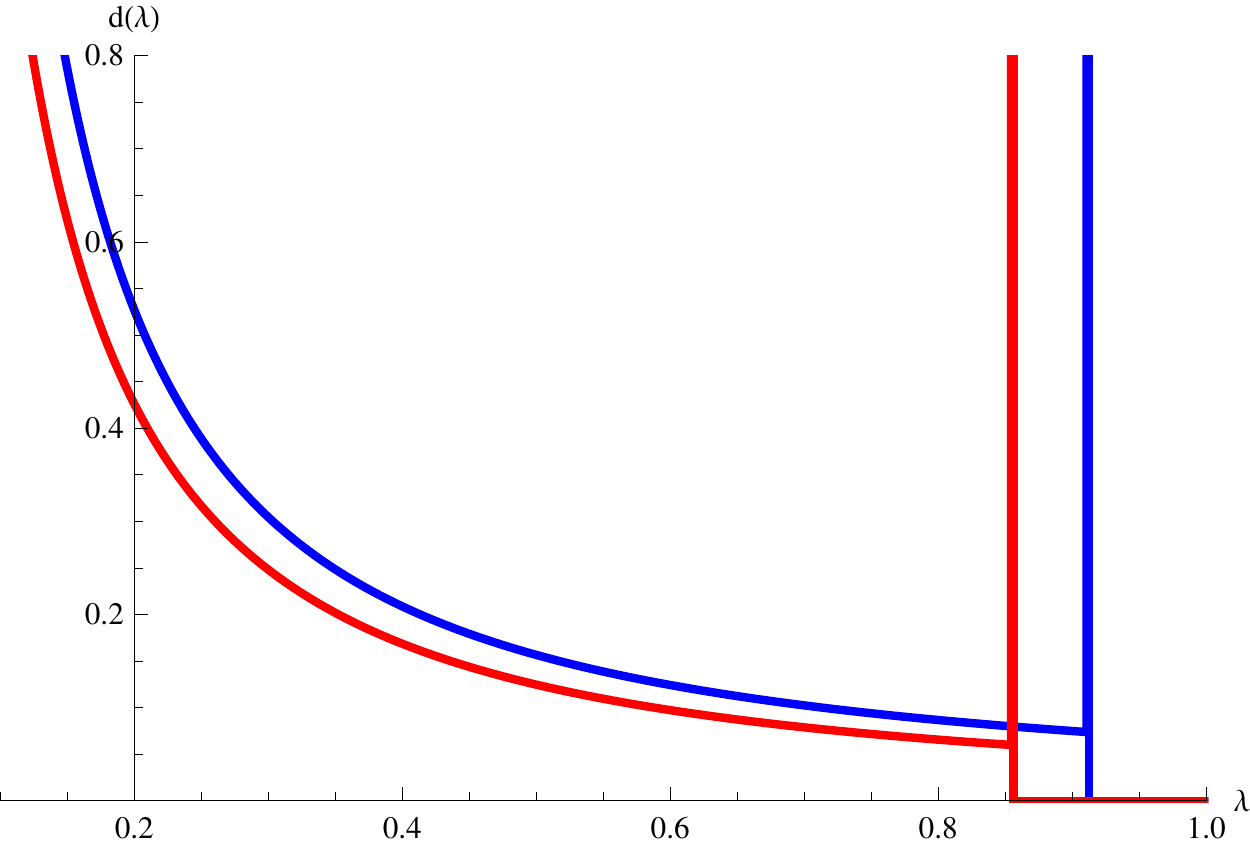}
\caption{The spectral function $d(\lambda)$ for the cases $\mu^2 L^2 = -1$, where the black hole is always Einstein, and $\mu^2 L^2 = -4$, where the black hole always has scalar hair. There are delta functions at the lowest eigenvalues, $\lambda^E_1=0.855$ for the Einstein (red line) and $\lambda^{ES}_1=0.911$ for the Einstein-scalar (blue line). For the pure Einstein case and with $V_H=G_N=L=1$, we obtain the same result as \cite{Hung:2011nu}.
}
\label{density-lambda}
\end{center}
\end{figure}

We can now go ahead and compute the spectral densities (\ref{reny-spectral}) explicitly.  To exhibit a simple numerical answer, we will set $V_{\H_{d-1}}=L=G_N=1$.
Performing an inverse Laplace transform numerically is a bit tricky, so we expand the left hand side of (\ref{reny-spectral}) in the large $n$ limit as
\beq
\exp\left((1-n)S_{n}\right)=\exp((s_0-s_1)-s_0 n)\left(1+{g_1\over n}+{g_2\over n^2}+{g_3\over n^3}+\cdots \right).
\eeq
where $S_n=\sum_{i=0}^{\infty} s_{i}n^{-i}$ and the $g_i$ are functions of the $s_i$.
The inverse Laplace transformation of the first term gives a delta function and the rest of the terms include the Heaviside step function and continuous functions. The constant $s_0$ determines both the location of the delta function and the 
Heaviside step function.
The result is 
\beq
d(\lambda)={\exp(s_0-s_1)\over \lambda}(l_0\delta(\lambda-\lambda_{1})+\Theta(\lambda-\lambda_1)\sum_{n=0} l_{n+1}(-\log \lambda)^{n}
)
\eeq
where $l_i$ is a function of $s_i$. Note that $-\log \lambda$ is the eigenvalue of the entanglement hamiltonian $H_E$ (\ref{modular hamiltonian}).
%
%
%\if0
%\bea
%d(\lambda)={1\over \lambda}\left(1.092\delta(t-0.156)+\Theta(t-0.156)(0.048+0.019t+0.002t^2+\mathcal{O}(t^3))\right) 
%\eea
%for the Einstein theory, and 
%\bea
%d(\lambda)={1\over \lambda}\left(0.994\delta(t-0.092)+\Theta(t-0.092)(0.066+0.016t+0.002t^2+\mathcal{O}(t^3))\right)
%\eea
%for the Einstein scalar theory. Here, $t=-\log \lambda$.
%\fi
%
%At $\mu^2L^2\ge -1$ and $\mu^2L^2= -4$, the critical temperatures are $T_c=0$ and $T_c=T_0=1/2\pi L$.
%In these cases, the integrations in (\ref{Rnyi-therm-2}) is either of  $S^{E}_{\text{therm}}(T)$ or  $S^{ES}_{\text{therm}}(T)$.
%We consider these two cases first.
%Since there is no phase transition between $T=0$ and $T=T_0$, the R{\'e}nyi entropy is a smooth function of $\al$ and it has a Taylor expansion 
%\beq
%S_{\alpha}=\sum_{n=0}s_{n}\al^{-n}
%\label{renyi-taylor}
%\eeq
The numerical result for the spectral density is plotted in Fig.\ref{density-lambda}. Of course, the R\'{e}nyi entropies are UV-divergent but we are calculating the constant in front of the divergent piece. Similar arguments can be found in section 5 of \cite{Hung:2011nu}.

It is worth noting that these results for the spectral density are consistent with those found by Calabrese and Lefevre  \cite{calabrese-Lefevre}.
In two dimensional conformal theory, the $n$ dependence of the R\'{e}nyi entropy is
\beq
S_{n}\propto \left(1+{1\over n}\right).
\eeq
This is determined by the conformal dimension of the twist operators.
By taking the inverse Laplace transformation, Calabrese and Lefevre obtained a universal scaling behaviour for 
the entanglement spectrum.
The structure of the spectrum, the delta function at the largest eigenvalue and the Heaviside step function, depends
only on the asymptotic behaviour of the R\'{e}nyi entropy at large $n$.  We have found a very similar structure for our 4d CFTs.
%However, the entanglement spectrum is not universal in $d>2$,  in that it depends on the cut-off. 
%%(We set the volume of the hyperbolic space to 1 in the computation of the black hole entropy and the following R\'{e}nyi entropy. 
%%As mentioned before, this should be replaced by the regularized volume of the hyperbolic space,
%%vol$(H_{d-1})L^{d-1}=\Omega_{d-2}\int_{1}^{L/\delta}(y^2-1)^{(d-3)/2}dy$, to see the cut-off $\delta$ dependence.) 
%It may be interesting is interesting to study the spectrum of the universal part of the entropy (sub-leading terms) or finite parts such as the mutual information in higher dimensional conformal field theories. We leave it for the future work.

For a general mass, it is technically difficult to perform the inverse Laplace transformation. 
However, from the asymptotic behaviour of the R{\'e}nyi entropy, we can see that the spectral function has only one delta function which sits between $\lambda^E_1$
and $\lambda_{1}^{ES}$.
Fig. \ref{dim-lambda1} shows the relation between the maximum eigenvalue of $H_E$ and $\Delta$.
Note that as we increase the dimension of the conformal operator $\Delta$, the maximum eigenvalue $\lambda_1$
of the entanglement spectrum monotonically decreases, i.e.
\bea
{ dS_{\infty}\over d\Delta}\ge0
\eea
This suggests that the ground state of a CFT with smaller lowest dimension operator
is closer to a pure state.  In other words, the ground state of a theory with a large gap is more mixed than one with a small gap.
This result is obtained using black hole thermodynamics, so is likely to be generic for large $N$ theories.
%In \cite{Bhattacharya:2012mi}, some analogy of the thermodynamic law for the entanglement entropy was investigated.
It would be interesting to investigate the generality of this inequality, and the analytic relation between $S_{\infty}$ and $\Delta$.

%%%%% begin Figure
%\begin{figure}[htb]
%\begin{center}
%\includegraphics[width=.45\textwidth]{cdensity.pdf}
%\caption{The spectral function $d(\lambda)$. There are discontinuous changes at $\lambda=0.855$ for the Einstein (pink line) and $\lambda=0.911$ for the Einstein-scalar (blue line).
%There are delta functions at those points, which are not shown in the figure.
%}
%\label{density-lambda}
%\end{center}
%\end{figure}
%%%%% end Figure

%%%%% begin Figure
%%%%% end Figure

%%%%% begin Figure

\section{Instability for the Normalizable Mode}% and a supercold phase transition 
\label{nonconstant}

We now turn our attention to the more physical case of the mode which is normalizable on the hyperboloid.  In this case the condensation of the mode will break the hyperbolic symmetries, so the construction of hairy black holes at the non-linear level is considerably more difficult.  We will simply perform the linearized analysis, which is sufficient to demonstrate that an instability exists and that the R\'{e}nyi entropies will undergo a phase transition. 
%The instability of these black holes is, to the best of our knowledge, new. 
We leave the construction of hairy black hole solutions to future work.
%
% and we leave that problem for future work (it would require changing the metric Ansatz of the hyperbolic directions, as the scalar field will break the symmetries of the hyperboloid; similar effects have been noticed in condensed matter theories cite[??]). 

%Without worrying about the end-point of the instabilities, we will perform a linear numerical analysis that shows that non-compact topological black holes are unstable at low temperature, when minimally coupled to a light scalar. This instability comes from the condensation of the normalizable modes of the scalar field on the hyperboloid and is responsible for a phase transition in the R\'{e}nyi entropies, as in the previous section

%However, this is only a statement about near-horizon geometry. In order to verify the validity of this argument, we need to analyse the wave equation, which is a non-linear function of the metric. 

In $d=4$, the wave equation for a scalar of mass $\mu$ is\footnote{Similar equations hold in other dimensions; we focus on $d=4$ for simplicity.}
\beq
\left(-\frac{\omega^2}{f(r)}-\frac{\lambda}{r^2}-\mu^2\right)\phi(r)+\left(f'(r)+3\frac{f(r)}{r}\right)\phi'(r)+f(r)\phi''(r)=0 \label{waveequation}
\eeq
where we considered the following ansatz for the field:
\bea
\Phi(t,r,\sigma_i)=e^{\omega t}\phi(r)Y(\sigma_i), ~~~\nabla^2_{\mathbb{H}_3} Y(\sigma_i)=-\lambda Y(\sigma_i).
\eea

The black hole will be unstable if (\ref{waveequation}) has a solution with $\omega$ real and positive with the field satisfying  specified boundary conditions at infinity and the horizon. %We will restrict our attention here to standard boundary conditions. 
We can put the wave equation in Schrodinger form by letting $\psi(r)=r^{(d-1)/2}\phi(r)$, so
\beq
\left(-\left(f(r)\frac{d}{dr}\right)^2+V(r)\right)\psi(r)=-\omega^2\psi(r) \label{schrodinger}
\eeq
with
\beq
V(r)=\frac{f(r)}{r^2}\left(\lambda+f'(r)\frac{d-1}{2}r+f(r)\frac{(d-1)(d-3)}{4}+\mu^2r^2\right)
\eeq
In tortoise coordinates ($dr_*=dr/f(r)$), this is the problem of determining whether the potential $V(r_*)$ has a negative energy bound state. 
We impose the following  boundary conditions \footnote{The field should vanish at the horizon as unstable modes behave as $\phi(r)\sim (r-r_+)^\omega$ as $r \to r_+$.}
\beq
\psi(r_+)=0 \ \ \ \ \ \ \ \ \psi'(r_+)=1 \ \ \ \ \ \ \ \ \ \psi(r)|_{r\to\infty}\to\frac{1}{r^{\Delta-(d-1)/2}} ~.
\eeq
Note that we consider here both Dirichlet and Neumann boundary conditions.

\begin{figure}[htb]
	\begin{minipage}[b]{0.47\linewidth}
     			\centering \includegraphics[scale=0.55]{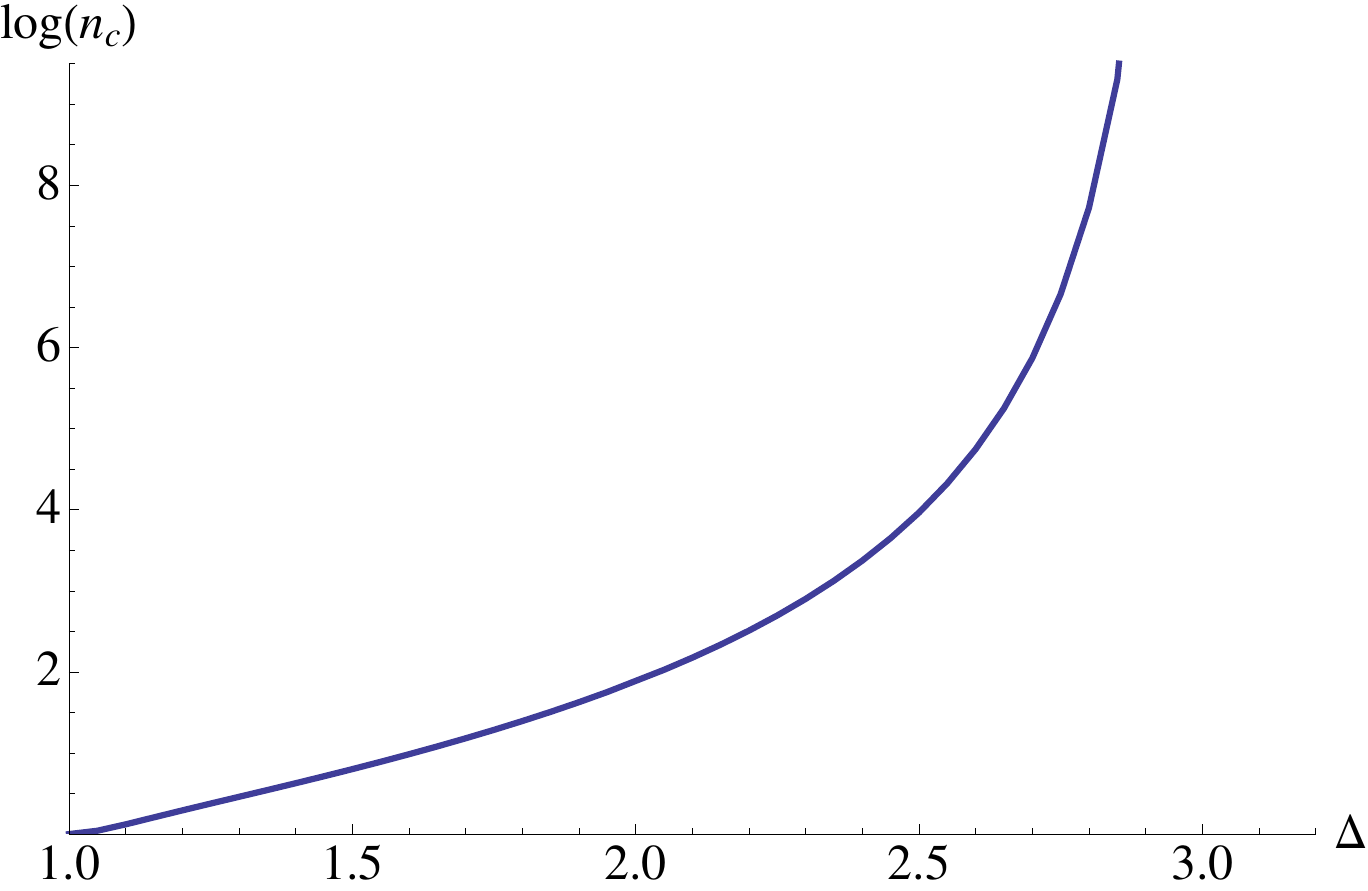}
      			\caption{Log of the critical R\'{e}nyi parameter for instability as a function of conformal dimension in $AdS_5/CFT_4$.  As $\Delta$ approaches $3$ the instability disappears.}
			\label{alpha_vs_delta}
	\end{minipage}\hfill
	\begin{minipage}[b]{0.47\linewidth}   
     			 \centering \includegraphics[scale=0.55]{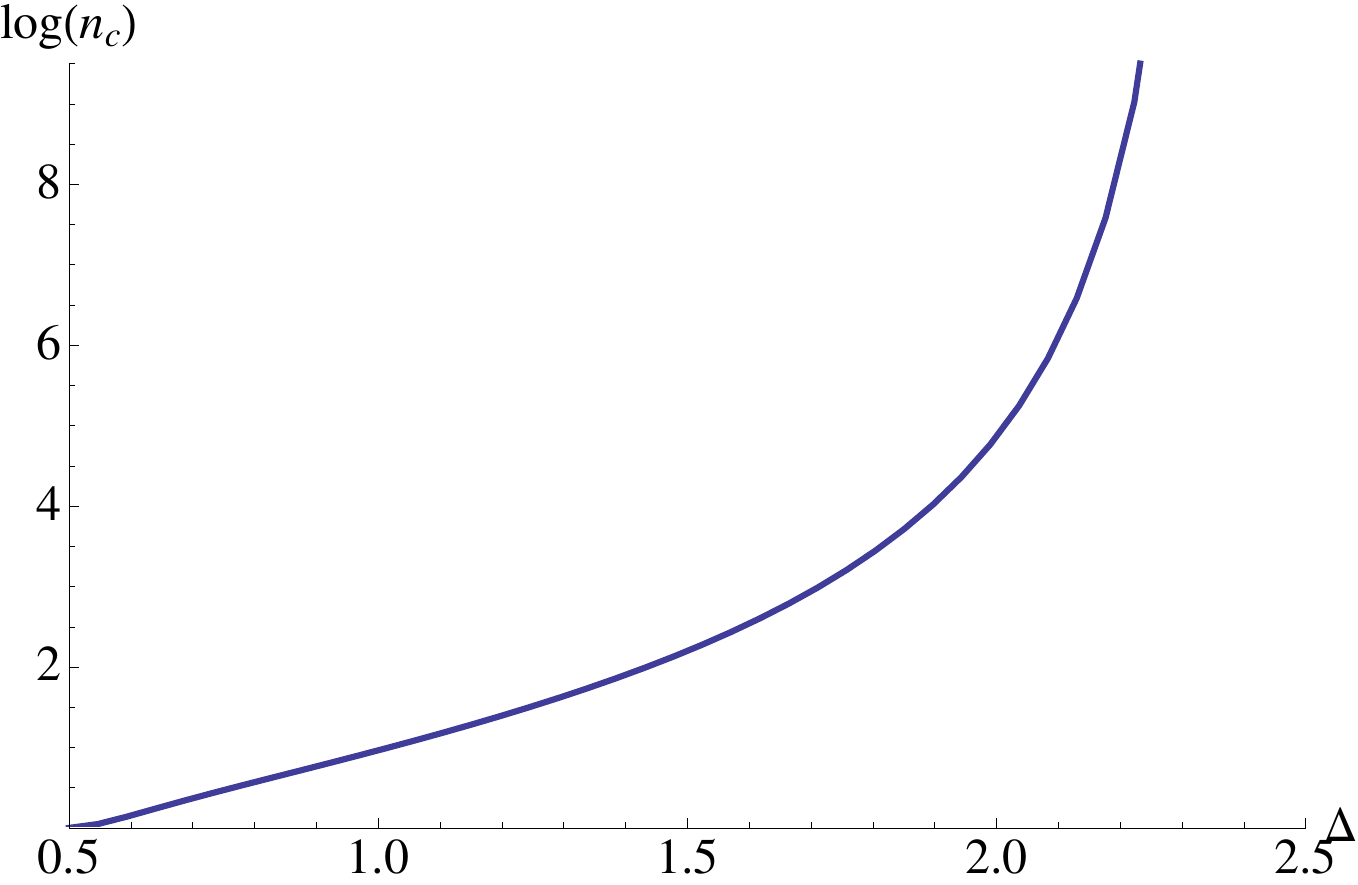}
     			 \caption{Log of the critical R\'{e}nyi parameter for instability as a function of conformal dimension in $AdS_4/CFT_3$. 
			 The instability disappears as $\Delta\to{3+\sqrt{3}\over 2}\approx 2.37$.}
			\label{alpha_vs_delta_d=4}
 	\end{minipage}  \hfill
\end{figure}

We can solve (\ref{schrodinger}) numerically using a standard shooting method from the horizon. 
We find that, for every mass in the range
\beq
-\frac{d^2}{4}\le \mu^2L^2_{AdS_{d+1}}\le-\frac{d(d-1)}{4} \label{massbound}
\eeq
there is a temperature $T >0$ for which the Einstein black hole is unstable to the development of scalar hair.\footnote{We have also checked this result using the method of trial wave functions, for which we thank G. Salton for useful discussions.}   
We have checked this in four and five dimensions.
%We will operate in the following way. %We fix the temperature of the black hole and find the mass for which the frequency is approaching zero, which will be the limit of instability. 
We plot the curve of marginal stability in Figs. \ref{alpha_vs_delta} and \ref{alpha_vs_delta_d=4} for four and five dimensions.  Every configuration in the zone above the curves is unstable.  
We conclude that  R\'{e}nyi entropies will undergo a phase transition at values of the R\'{e}nyi parameters $n_{c}$ which depend on the dimension of the lowest nontrivial operator.  
 
%%%%% Begin Figure

%%%%% End Figure

These results have one important feature which distinguishes them from the instabilities considered in the previous section.  They take place at $much$ lower temperatures (i.e. larger values of $n$).  It turns out that, for scalars which are non-constant on the hyperboloid, the $AdS_2\times \H_{d-1}$ instability becomes relevant only very close to extremality.  A similar phenomenon was noted in \cite{Dias:2010ma} for a different class of black holes.  This suggests that it is primarily the lowest eigenvalues of the entanglement spectrum which are effected by the scalar field instability.

\section{Acknowledgements}

We thank 
Jaume Gomis,
Nabil Iqbal,
Kristan Jensen,
Max Metlits,  
Rob Myers,
Harvey Reall,
Subir Sachdev,
Jorge Santos,
Stephen Shenker,
and
Tadashi Takayanagi
for helpful discussions.
This research was supported by the National Science and Engineering Research Council of Canada.
SM thanks 
University of British Columbia, University of Victoria, Kavli Institute for the Physics and Mathematics of the Universe (Kavli IPMU), Kyoto University and the Perimeter Institute for theoretical physics for hospitality.

%%%%%%%%%%%%%%%%%%%%%%%%%%%%%%%%%%%%%%%%

\end{document}